\documentclass[conference]{IEEEtran}
\IEEEoverridecommandlockouts

\usepackage{cite}
\usepackage{amsmath,amssymb,amsfonts}
\usepackage{algorithmic}
\usepackage{graphicx}
\usepackage{textcomp}
\usepackage[normalem]{ulem}
\def\BibTeX{{\rm B\kern-.05em{\sc i\kern-.025em b}\kern-.08em
    T\kern-.1667em\lower.7ex\hbox{E}\kern-.125emX}}
\usepackage{xurl}

\newcommand{\dace}[0]{\mbox{DaCe}}
\newcommand{\sdfg}[0]{\mbox{SDFG}}
\newcommand{\ftodace}[0]{\mbox{F2DaCe}}

\newcommand{\gtforpy}[0]{\mbox{GT4Py}}
\newcommand{\kokkos}[0]{\mbox{Kokkos}}
\newcommand{\openmp}[0]{\mbox{OpenMP}}
\newcommand{\openacc}[0]{\mbox{OpenACC}}
\newcommand{\stdpar}[0]{\mbox{C++ \texttt{std::par}}}
\newcommand{\sycl}[0]{\mbox{SYCL}}

\newcommand{\mask}[0]{\textsc{Mask}}
\newcommand{\sig}[0]{\textsc{SIG}}
\newcommand{\flip}[0]{\textsc{Flip}}

\usepackage{fontawesome5}
\usepackage{units}
\usepackage{hyperref} 
\usepackage[capitalize]{cleveref}
\usepackage{soul} 
\soulregister\cite7
\soulregister\ref7
\soulregister\cref7
\soulregister\pageref7
\crefname{figure}{Figure}{Figures}
\crefname{appendix}{Appendix}{Appendices}

\usepackage{mathtools}  
\usepackage{xspace}     

\usepackage{booktabs, tabularx}
\newcolumntype{R}{>{\raggedleft\arraybackslash}X}
\usepackage{multirow}
\usepackage{colortbl}
                                   
\newcommand{\rC}[0]{\rowcolor[HTML]{EEEEEE}}                                    
\newcommand{\hC}[0]{\rowcolor[HTML]{333333}}

\usepackage[dvipsnames]{xcolor}

\definecolor{hlgray}{gray}{0.9}
\sethlcolor{hlgray}

\definecolor{rqblue}{RGB}{220,235,252}
\newcommand{\researchquestion}[1]{%
    \smallskip\noindent\fcolorbox{black}{rqblue}{\parbox{\dimexpr\linewidth-2\fboxsep-2\fboxrule}{\emph{#1}}}\smallskip}
\newcommand{\hypothesis}[1]{%
    \smallskip\noindent\fcolorbox{black}{rqblue}{\parbox{\dimexpr\linewidth-2\fboxsep-2\fboxrule}{\emph{Hypothesis}: #1}}\smallskip}

\begin{document}

\title{Performance Portability Explained: \\ Insights from accelerating ICON microphysics
}

\author{

\IEEEauthorblockN{
Faveo H\"orold\textsuperscript{1,2,*},
Georgiana Mania\textsuperscript{3,*}
}
\IEEEauthorblockN{
Panagiotis Adamidis\textsuperscript{3},
Harshada Balasubramanian\textsuperscript{4},
Jörg Behrens\textsuperscript{3},
Yakup Budanaz\textsuperscript{1},
Alexandru Calotoiu\textsuperscript{1},
}
\IEEEauthorblockN{
Nicoletta Farabullini\textsuperscript{5},
Andreas Herten\textsuperscript{2},
Torsten Hoefler\textsuperscript{1,5},
Yannik Koenneker,
Ioannis Magkanaris\textsuperscript{5}
}
\IEEEauthorblockN{
Philip Müller\textsuperscript{5},
Ralf Müller\textsuperscript{3},
Edoardo Paone\textsuperscript{5},
William Sawyer\textsuperscript{5},
Bjorn Stevens\textsuperscript{4},
Hannes Vogt\textsuperscript{5}
}
\\
\IEEEauthorblockA{
\textsuperscript{1}\textit{ETH}, Zurich, Switzerland \\
\textsuperscript{2}\textit{Forschungszentrum J\"ulich}, J\"ulich, Germany \\
\textsuperscript{3}\textit{Deutsches Klimarechenzentrum (DKRZ)}, Hamburg, Germany \\
\textsuperscript{4}\textit{Max Planck Institute for Meteorology}, Hamburg, Germany \\
\textsuperscript{5}\textit{Swiss National Computing Centre (CSCS)}, Zurich, Switzerland
}

\thanks{* co-lead, equal contribution. Email: fhoerold@ethz.ch}

}


\maketitle

\begin{abstract}
Scientific computing is under increasing pressure to adapt to HPC hardware that
is shaped by AI workloads. In these modern ecosystems, extracting reasonable
performance without investing major engineering effort is a central issue for
domain scientists. Correctness constraints and intricate reference
implementations represent further impediments to development.  Instead of using
a double-precision implementation as a reference, considering the underlying
scientific problem directly opens more opportunities for performance
improvements when updating and rewriting applications.

We take seven different portability frameworks---covering pragma-based, DSL,
and data-centric approaches---and evaluate them on a microphysics kernel of the
ICON climate model. Our performance benchmarks show that each different HPC
system benefits from a different combination of optimizations, so we develop the
measure of framework expressivity to gauge how easily each framework can
customize portable implementations in this way.
\end{abstract}



\section{Introduction} \label{sec:intro}

It is imperative for scientific computing in the HPC context to adapt to hardware ecosystems that are increasingly shaped by AI  workloads. In weather and climate modeling, this shift is occurring alongside a broader transformation in both the scientific ambition and computational organization of Earth-system simulations, as kilometer-scale models and ensemble-based workflows demand ever greater computational efficiency~\cite{bauer21,bauer15}.
Not only are scientific applications expected to be performance portable even on
the newest exascale systems, but results must remain correct across different
hard- and software environments ~\cite{shalf20}.
%
%
Evaluating correctness of new implementations is difficult, regardless of whether
results are within tolerance for a certain reference, or physically correct for a set of standard experiments with known solutions ~\cite{operkampf02, rudisuhli14, duben17, gunsteren18, banderier23}.



In general, optimizing and porting scientific applications to modern accelerator
architectures does not result in a faithful re-expression of an existing
implementation, 
especially given recent trends that change floating-point arithmetic at the hardware level. Furthermore, for many real-world problems there is no reference solution to compare against, only a reference implementation which has shown to match expectations for a set of experiments empirically. Approximating the difference between the physical solution and a given implementation is a research field of its own.
Instead, performance portability requires design choices at multiple levels
for the efficient implementation of computations and, in tandem, the representation of the underlying problem in terms of physics and numerics.

Here, we argue that a scientific code development cannot be defined through a sequence of
discrete, independent decisions. Rather, the best approach is for scientists to
jointly navigate the shared design space with a spectrum
of abstraction levels covering physics, mathematics, and computer science.
From the precise form of equations employed, to the applied numerical algorithms and
discretization strategies, down to the order of operations chosen at the
hardware level, each decision has ripple effects on the final application.
The \textit{domain scientist} generally makes decisions about physical
models and the experiment setup at the highest level of the spectrum.
The \textit{computational scientist} bridges the gap between the domain scientist and the performance engineer by applying existing numerical methods to discretize the model ~\cite{sameh96,cunningham02}. 
At the
lowest level, the \textit{performance engineer} uses her knowledge of computer science
to introduce efficient optimizations. In a productive workflow, all three
scientists will be in active communication  and aware of the broad
impact that decisions at any level of the design space can have.
In order to demonstrate how a simple design decision at the low-level
end of the design space can affect all levels, we run performance portability
benchmarks with a small physics kernel from the weather and climate domain,
called \textit{Graupel}.
Like many scientific codes, the Graupel kernel is sensitive to perturbations in
computations, which arise from the fact that addition and multiplication are not
associative in floating-point arithmetic.
While $a+b+c+d$ may be a mathematically unambiguous equation to the domain scientist, the computational scientist may prescribe a specific associativity $(((a+b)+c)+d)$ to best avoid numerical cancellation.
At an even lower level, the performance engineer may prefer a different
associativity $((a+b)+(c+d))$ in order to utilize superscalar hardware more
efficiently.
Thus, the shift in semantics at a lower level may undermine design decisions that
have already been made at a higher level.

In this work, we use the Graupel kernel to argue that achieving performance on
modern architectures is not simply a matter of re-expressing an existing
implementation in a portable framework. Rather, the rewriting process
frequently traverses several levels of abstraction, as implementation choices
interact with numerical structure, latent assumptions in the reference code, and
the acceptable interpretation of the underlying scientific problem. Different
performance portability frameworks mediate this process in different ways: they
expose different optimization opportunities, impose different constraints on the
resulting implementation, and support different modes of collaboration between
domain scientists and performance engineers.

We accelerate Graupel using seven different frameworks: \openmp{}~\cite{openmp4}, \openacc{}~\cite{OpenACC34}, \kokkos{}~\cite{trott22},
\sycl{}~\cite{10.1145/3388333.3388649}, \stdpar{}, \dace{}, and \gtforpy{} \cite{paredes2023gt4pyhighperformancestencils}. The Graupel one-moment
microphysics module is part of the ICON climate and weather simulation
ecosystem, which has been the focus of an ongoing effort to enable efficient
kilometer-scale simulations on exascale HPC systems~\cite{adamidis25}.
Graupel, and by extension ICON,
represents the perfect proving ground for testing our ideas about performance
portability.
The following contributions are made:
\begin{itemize}
\item{We apply seven different parallelization, performance, and portability frameworks to Graupel and benchmark the resulting implementations on several HPC systems.}
\item{After identifying the optimal set of optimizations on each system, we discuss how the engineering effort required to port, parallelize, and optimize Graupel varies between frameworks.}
\item{We distill the overlapping features of each framework down to a measure of \textit{framework expressivity}, which evaluates how easily each framework can produce the distinct sets of optimizations needed for the different HPC systems.}
\end{itemize}

Moving forward, we introduce ICON and Graupel in \cref{sec:icon}.
Next, we explore the design decision spectrum using concrete examples from
Graupel in \cref{sec:spectrum}.
Then, we introduce the seven frameworks in \cref{sec:frameworks}, and present
our performance results in \cref{sec:results}, where we also field test our
implementations in a full-scale distributed R2B8 ICON run.
In \cref{sec:expressivity}, we introduce the concept of \textit{framework
expressivity} and examine how it can reduce engineering efforts.
In \cref{sec:related}, we summarize related frameworks and move to a conclusion in \cref{sec:conclusion}.

\section{ICON and Graupel one-moment microphysics} \label{sec:icon}

ICON (ICOsahedral Non-hydrostatic) \cite{ICON} is a code base that supports a variety of atmospheric and oceanic applications, mainly related to the domains of numerical weather prediction and climate modeling.  All atmospheric applications share an underlying fluid dynamical solver, which can be configured for various geometries \cite{fievet26}, e.g., on an icosahedral grid \cite{zaengl15}, and resolutions \cite{klocke25}, including static adaptivity through nesting. The dynamical solver includes an energy equation coupled to mass conservation equations. Additionally, a transport module advects various phases and habits of water, as well as trace-gases or particulates. Diabatic processes, such as turbulence, radiation, or particulate microphysical processes are also modeled.  

The ICON code base contains different cloud-microphysical models, but the core
approach is a one-moment bulk scheme called \textit{Graupel} that  describes the
formation, growth, and sedimentation of water particles (hydro\-meteors).
\textit{Graupel} calculates transitions between vapor, liquid and solid water,
whereby the liquid and solid phases are described by different habits---
droplets and drops (rain) for liquid; ice, snow, and graupel for solid. Phase
changes influence temperature and density. This in turn influences force balances in
the dynamic equations both directly, and indirectly by virtue of how water is
partitioned among phases and habits, and how these interact differently with radiant
energy transfer.

Graupel presented itself as an interesting case study for the study of
performance portable design choices for two main reasons: Firstly, 
\textit{Graupel} only involves grid dependencies in the vertical dimension,
where all points are co-located on a process. No message-passing communication
or horizontal reductions are required. This is an access pattern shared by most physical parameterizations in the weather
and climate domains. Secondly, \textit{Graupel} employs a sparse calculation
in physical space, reflecting that clouds and hydrometeors are not everywhere:
typically they cover only 5-25 percent of the domain. Making
it more challenging is that this sparsity is dynamic and threshold dependent.
To minimize computation on classical CPU architectures, 
various gather-scatter approaches are applied to condensate thresholds that
themselves could be targets of optimization.  For emerging architectures,
however, a gather-scatter approach is not necessarily advantageous. With this in
mind, alternative implementations only need to reproduce the results of the
masked domain, with some flexibility in how conservatively to define the
masks, mindful of the fact that the sparsity of this algorithm is the key
challenge in creating performance-portable code.
We take a closer look
at these sparsity-aware optimizations and examine the actual effects that they
have on performance for GPU architectures in
\cref{subsec:sparsity}.

As a first step in this study, the original ICON
 \textit{Graupel} implementation (still in ICON) for weather prediction was completely rewritten, still in Fortran, but avoiding advanced Fortran language features. The rewrite was done by a domain scientist, with the intent of removing unnecessary dependencies and unused legacy functionality, and  separating specific processes that had become entangled over years of piecewise optimization. The rewrite resulted in a threefold reduction in lines of code, a much greater algorithmic transparency with no loss of efficiency on classic CPU architectures, and the elimination of some errors.  This rewrite made the compact code base more accessible for computational scientists and performance engineers for subsequent analysis.


In addition to the sequential Fortran implementation, a parallelized version of the Graupel code using \openacc{} is available in the official ICON release. Since Graupel does not make use of any distributed reductions, optimizations can produce bit-identical results when built with a conservative compiler configuration.
Nevertheless, those optimizations that adjust the handling of sparsity for accelerators fundamentally change the numerical results of Graupel. The necessity of discussing, handling, and accepting such changes is one of the main discussion points of this work. 


\section{A spectrum of design choices} \label{sec:spectrum}


\begin{figure}
  \centering
  \includegraphics[width=\linewidth]{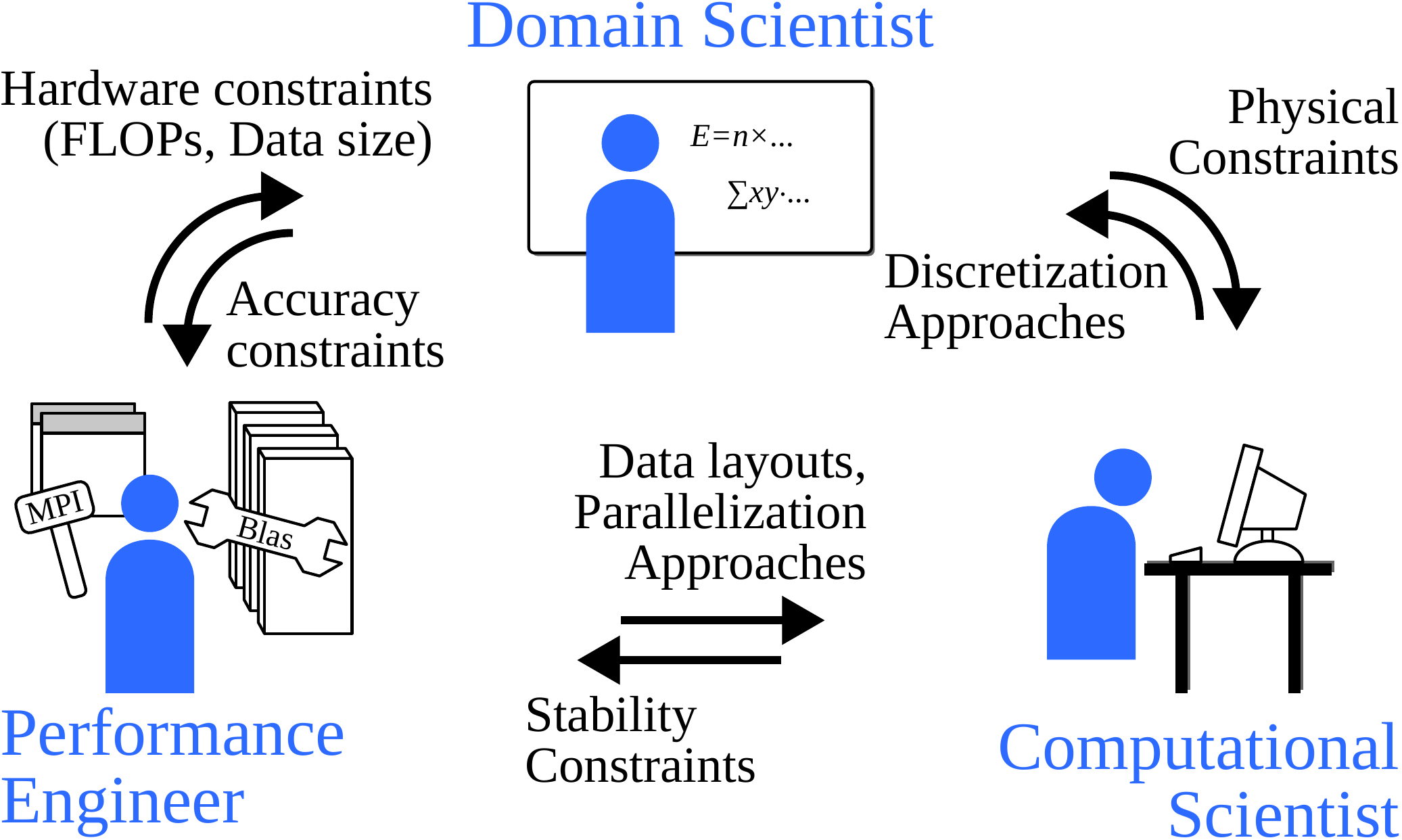}
  \caption{
     Illustrating typical design choices at abstraction levels and coordination
     between experts. At the highest level, the domain scientist formulates a
     hypothesis and chooses the equations which model her problem. At an
     intermediate level, the computational scientist discretizes physical and
     temporal domains, introducing appropriate numerical methods to solve the
     scientist's equations. At the lowest level, the performance engineer
     chooses data layouts and parallelization strategies to map the numerical
     methods to a given HPC architecture.
  }
  \label{fig:designers}
\end{figure}


In this section, we build on our ideas from \cref{sec:intro}
and solidify how all design choices that go into designing, porting, and
optimizing a scientific application can be mapped onto a spectrum passing
through each of the domain scientist's, computational scientist's, and
performance engineer's domains.

The highest levels of the design space reside in the domain of physics. This is
where the domain scientist comes up with a hypothesis, constructs a physical
model, and outlines an experiment to prove or disprove the hypothesis. The
domain scientist is responsible for setting initial and boundary conditions for
the experiment, accuracy constraints, and physical equations to simulate.

The layers at a medium level of abstraction in the design space reside in the
space of mathematics. Here, the computational scientist takes the experiment
proposed by the physicist, discretizes it, and chooses the appropriate numerical
methods to solve the physical equations while fulfilling the accuracy
constraints imposed by the physicist. The computational scientist is responsible
for imposing stability constraints, which will sometimes imply floating point
accuracy constraints.

The lowest abstraction levels of the design space dive into computer science.
Here, the performance engineer will take the high-level implementation proposed
by the computational scientist and look for opportunities to improve scheduling,
data layout, and access patterns. They have to work under the constraint of
the stability and accuracy constraints imposed by the domain and computational
scientists, respectively. However, they can also ask the computational scientist
to update their numerical model, which might be made more amenable to efficient
data layouts if the discretization approach (e.g. grid, domain decomposition) is
adjusted. The performance engineer's feedback can even influence decisions about
the experiment design at the highest abstraction levels; depending on hardware
constraints like available FLOPs, memory, and input/output data storage, the
physicist may need to update their experiment and the computational scientist
may need to find more efficient numerical methods or propose a more granular
domain decomposition to make better use of limited HPC node resources.

In practice, no single person in a research team will have a deep understanding
of every level in the abstraction spectrum. Rather, various scientists
specialize in different regions of the spectrum and must communicate in order to
choose the best approach that leads to a feasible design at the low level while
delivering acceptable results at the high level.

Across all of these levels, design choices may be semantically equivalent within
the abstraction at which they are made, while nevertheless constraining the
choices that remain available at lower levels. Crucially, once a decision has
been fixed at a higher level, subsequent transformations at lower levels may be
unable to recover either the numerical behavior or the performance potential
that would have been enabled by a different choice at the higher level. At the
same time optimizations such as reordering instructions at the lower level
cannot be applied without fundamentally breaking numerical approaches at the
higher level. These inter-dependencies and the need for coordination between
experts is illustrated in \cref{fig:designers}.

When confronted with the need to rewrite or port an application to new hardware, the
central question is no longer merely how to preserve an existing implementation,
but how far one must ascend the ladder of abstraction in order to recover a
sufficiently rich optimization space. The answer will, in general, depend on
both the structure of the application and the characteristics of the target
system and the definition of correctness.

CLEO\cite{bayleynumerical} illustrates the cumulative effect of design choices
at different abstraction levels. It is a state-of-the-art cloud microphysics
model in ICON that uses Lagrangian particle tracking to explicitly compute
microphysical processes, thereby providing a potential alternative to the
reference implementation used in this work. With the introduction of a more
computationally intensive physical model by the domain
scientist\cite{bayleyhpc}, both the computational scientist and the performance
engineer focus on designing for and improving flexibility and performance,
instead of merely optimizing an existing model.

In this work, we focus on the central part of the design spectrum, encompassing
implementation details and optimizations related to the HPC and node-local
architectures. The design decisions required for parallelization and porting
also fall within this region. Therefore, we explicitly do not consider the edges
of the design spectrum: In our situation, the domain scientist has already fixed
the physical model and the computational scientist has already decided on a
numerical scheme and discretization strategy. However, we conferred with domain
experts to confirm that propagating or masking out certain negligible physical
contributions was acceptable for the results of the Graupel simulation. We
explore the effects of these sparsity-exploiting optimizations in
\cref{subsec:sparsity}. On the other end of the spectrum, details of the ISA and
hardware implementation have been fixed; we run our computations using IEEE-754
single- and double-precision floating-point arithmetic on various x86-64 and
aarch64 processors, optionally with offloading to Nvidia and AMD GPU
accelerators.

\section{Frameworks for Parallelization, Performance, and Portability} \label{sec:frameworks}





This section introduces the frameworks and APIs that are currently being evaluated, roughly ordered by historical appearance: \openmp{}, \openacc{}, \kokkos{}, \sycl{}, C++ Standard Parallelism (\stdpar{}), \dace{} (using its Fortran frontend) and \gtforpy{} (using \dace{} as a performance-enhancing backend). All frameworks aim to allow the same source code to target multiple hardware architectures while aiming for good performance during parallel execution, but choose different means to achieve this goal.  
Support of a certain programming model on a certain device type through a certain compiler is a complex and moving target. While we try to give a concise overview, we refer to the literature for more extensive discussions~\cite{manygpumodels}.

Despite the rewrites being largely independent, there are overall trends that
can be identified across the different frameworks:
In contrast to directive-based approaches, \kokkos{}, \sycl{}, and \stdpar{} do
not merely annotate an existing sequential implementation. Instead, they require
developers to reformulate the computation in terms of framework abstractions.
As a result, although the host language remains C++, achieving portability is
not simply a matter of incremental annotation, but of recasting the algorithm so
that parallelism is made explicit in the form expected by the framework. 
The resulting code is shaped not only by the scientific algorithm itself but
also by the abstractions and limitations imposed by the respective frameworks.
This can make implementations more challenging to read and reason about,
particularly for computational and domain scientists whose primary concern is
the numerical method rather than the mechanics of heterogeneous execution. On
the other hand, directive-based frameworks like \openmp{} and \openacc{} may
require compile-time logic to differentiate between different acceleration
approaches, which is also known to clutter code and make maintenance more
challenging. Finally, \dace{} and \gtforpy{} side-step this issue by hiding
optimizations from the code altogether. The potential downside is that the build
process for optimized release binaries becomes more complex.

We note that the initial translation of the Fortran code to sequential C++ using
raw pointers served as the starting point for \openacc{}, \openmp{}, \sycl{},
and \stdpar{} implementations, hence they share some of the technical decisions,
i.e., to use linearized array indexing. While \kokkos{}, \dace{} and \gtforpy{}
make use of linearized arrays in the backend/generated code, the 2d indexing is
still used in the application code, which improves readability.




\subsection{\openmp{}} 
Introduced in 1997, \openmp{} is arguably the most prominent parallel programming model for shared- and near-memory parallelism~ \cite{kadosh23}.
At its core, the \openmp{} standard outlines an extensive set of compiler directives designed to facilitate parallel programming. These directives require compiler support for translation into parallel constructs, alongside a specialized runtime library that implements hardware-specific functionality.

The \openmp{} standard v4.0 introduced the \texttt{omp target} directive (extended with every subsequent release), which opens the path to offloading the computations to accelerators such as GPUs.~\cite{openmp4} Although C++ and Fortran compilers provide well-optimized \openmp{} support for CPU architectures, GPU implementations still lag in both functional capabilities and performance efficiency.~\cite{jarmusch24} Hardware vendors develop specialized \openmp{} implementations, finely tuned to their architectures. NVIDIA’s HPC SDK compiler suite stands out as a notable example. Complementing this, open-source compilers like GNU GCC and LLVM Clang also offer strong \openmp{} support,~\cite{huber22} fostering accessibility and community-driven innovation.

The Graupel implementation using \openmp{} was tailored for GPU execution and in
view of exascale systems. Specific optimizations to reduce branching were
implemented alongside a series of algorithmic changes. Nevertheless, these
proved to hinder the CPU parallelism. This highlights a significant issue with
pragma-based portability frameworks---it is challenging to fine-tune the
implementation for heterogeneous set of architectures such as CPU and GPU
without sacrificing readability and maintainability.

\subsection{\openacc{}} 
\openacc{}~\cite{OpenACC34} is a directive-based programming model that targets
parallel accelerators. 
Supported mainly by NVIDIA, coverage beyond the NVIDIA HPC SDK compilers is limited, especially for CPUs.
\openacc{} is modeled closely after \openmp{}, but specifically for more detached accelerators (although CPU support is also available). In a time, when GPUs rapidly entered the HPC market, \openacc{} was able to swiftly develop directive-based accelerator programming, and implement it in vendor compilers. Simple directives, like \texttt{kernel} allow to accelerate many lines of code at once. As support for the model beyond the NVIDIA ecosystem is minimal, other mature accelerator-enabled alternatives have emerged since.

The Graupel C++ \openacc{} implementation is straight-forward and based on the sequential C++ version. Given the limited compiler support, no further optimizations were made, since we consider that C++ \openacc{} is not a viable successor of the existing Fortran \openacc{} in ICON.

\subsection{\kokkos{}} \label{subsec:kokkos}
\kokkos{}
 originated as part of the Exascale Computing Project
and split off into an independent project in 2015.
Currently, it is organized as an \textit{established} open-source project of the High Performance Software Foundation\footnote{\url{https://hpsf.io}}.

With powerful abstractions for data and execution modes, \kokkos{} enables scientists to separate the concerns between algorithm development and performance portability. The API is exposed as a C++ header-only library which translates the code to parallelization APIs native to each hardware platform, enabled by selecting the host and device backend at compile time. Currently, \kokkos{} supports three execution models for the host---a sequential one and two parallel ones, \openmp{} and C++ Threads---and three execution models for accelerators, enabled by the device backends CUDA, HIP, and \sycl{}. \kokkos{} is designed to be compiler-agnostic and it can therefore be effortlessly embedded into diverse software build systems.

The \texttt{Kokkos::View} abstraction for multi-dimensional arrays draws a conceptual parallel to Fortran’s array notation, easing ICON developers’ transition from Fortran to \kokkos{}. It also serves as the foundational blueprint for the upcoming \texttt{std::mdspan} support in the C++26 standard.

The \kokkos{} Graupel implementation is a direct rewrite of the original Fortran code, with minimal changes, which preserves the initial algorithm  structure for readability purposes. Therefore, it can trivially produce bit-identical results to Fortran in specific compilation setups on CPU platforms, while leaving room for future more aggressive optimizations.

\subsection{\sycl{}} 

The \sycl{} standard was introduced by the Khronos Group in 2014 as an extension to C++, aiming to build on the device support and parallel execution capabilities of OpenCL. \sycl{}, much like \kokkos{}, occupies an intermediate position in the performance-portability landscape: it preserves a single-source programming model across heterogeneous architectures, while still exposing enough structure for downstream compilers and runtimes to specialize execution for a given target.

Like \kokkos{}, \sycl{} has its own abstractions for execution and data management, imposing similar constraints. Additionally, unlike \kokkos{}, \sycl{} relies on compiler support, which provides added benefits such as \sycl{}-specific optimisations and just-in-time compilation for its C++ kernels, offering the flexibility of on-the-fly code generation for specific hardware---but of course at the cost of a strong compiler dependency.

The microphysics implementation in \sycl{}, a direct translation of the original Fortran code, was tested using Intel's oneAPI Data Parallel C++ (DPC++) compiler suite and AdaptiveCpp, which leverages LLVM to compile \sycl{} across different hardware platforms. On x86 hosts, the Intel compiler supports CUDA, ROCm/HIP, and Level Zero backends. Even so, separate plugins (e.g., for CUDA and ROCm/HIP) must be compiled and patched into the compiler manually, and are not always available on existing HPC systems. Similarly, AdaptiveCpp almost always needs to be built by the performance engineer, requiring additional effort in existing portability workflows. Moreover, on Arm CPUs, a heavy reliance on AdaptiveCpp as the single \sycl{} implementation remains a limiting factor.

\subsection{\stdpar{}} \label{subsec:stdpar} 

C++ Standard Parallelism, or \stdpar{}, refers to the support of parallel
execution directly in the C++ standard, enabled through the STL. Similar to
\kokkos{} and \sycl{}, individual algorithmic building blocks, like
\texttt{for\_each}, can be used to express algorithmic patterns abstract on a
high level. With the according execution policy, enabled algorithms can then be
executed in parallel and \textit{unsequenced}\footnote{Function calls with the
\textit{unsequenced} may be interleaved.}. \stdpar{} needs compiler and library
support, and is currently available in the NVIDIA HPC SDK or in the open-source
AdaptiveCpp.

The simplicity of clean C++ code—free from pragmas and explicit memory
control—comes with a limitation: STL parallel loop lambdas can only capture host
pointers, by value or reference. While this is not a problem for CPU runs, it
becomes one for GPU execution. In ICON’s use case, data pointers are copied to
the GPU before the time loop, processed across multiple timesteps, and copied
back to the host at the end. This necessitates passing device pointers directly
to the Graupel function to avoid CPU-GPU intermediate transfers. However, this
is not supported by \stdpar{}, leading to the high runtimes observed in
Figure~\ref{fig:gpu_runtimes} due to host-device memory transfers at every
invocation.



\subsection{\dace{}} \label{subsec:dace}


\dace{} is a data centric
framework~\cite{bennun19,ziogas21,bennun22,rausch22,boudaoud25,schaad25} that
provides architecture-specific backends, which specialize a common intermediate
representation---the Stateful Dataflow multiGraphs (\sdfg{}) IR---to diverse target systems, including modern CPUs
and GPUs. Its core contribution is to lift programs from control-centric
languages such as Fortran and Python into the data-centric representation, where
data movement and dependencies become explicit and latent parallelism can be
exposed before the computation is specialized toward a particular hardware
architecture. This separation is important: rather than embedding optimization
decisions directly into the source code, \dace{} enables transformations to be
applied at the level of the representation, allowing the same computational
intent to be retargeted across CPUs and accelerators. In this sense, \dace{} does
not merely offer another portability layer; it offers a structured pathway from
legacy scientific code to architecture-aware implementations through a sequence
of semantically equivalent program transformations in the performance engineering space.

In particular, data-centric
methods have been shown to support Gordon Bell Prize-winning work at
SC19,~\cite{ziogas19} and the same broader line of work is now also associated
with the SC25 Gordon Bell Prize for Climate through its role in enabling
efficient kilometer-scale Earth-system simulations~\cite{klocke25}. 


\dace{}'s vision is to separate the concerns of domain
scientists and performance engineers. Domain scientists should be able to
express the problem to be solved in a form that remains close to the underlying
scientific intent, while performance engineers operate on transformations that
improve performance, exploit parallelism, and specialize the computation for
different hardware targets. This division of labor improves maintainability and
readability and avoids forcing either group to acquire deep expertise in the
other’s discipline. As architectures
increasingly privilege mixed precision, heterogeneous execution, and AI-oriented
computational patterns, successful optimization cannot always remain confined to
backend tuning alone. In many cases, achieving substantial gains will require
revisiting the representation of the problem itself and, in some instances,
reconsidering aspects of the numerical method. 
While \dace{} allows the source code to stay in its original format, and thus
makes collaboration with the domain scientist and computational scientist
simpler, it does require the performance engineer to become familiar with the
data-centric representation and learn how to write and apply data-centric code
transformations, which is a significant challenge, different than optimizing
imperative programming languages, further compounded by the currently growing,
but small, community of experts. 

Here, we use \ftodace{} to refer to the \dace{} Fortran frontend, which lifts
the Fortran source code directly into the \sdfg{} IR.




\subsection{\gtforpy{}} 
\gtforpy{} \cite{paredes2023gt4pyhighperformancestencils} is a Python library for generating high performance implementations of stencil kernels from a high-level definition using Python syntax by taking advantage of a range of backends to provide high performance on a variety of compute architectures. It follows a functional programming model, providing abstractions for parallel patterns common in atmospheric simulation, and is part of the GridTools \cite{AFANASYEV2021100707} framework, a set of libraries and utilities used to develop performance portable applications for weather and climate modeling. Similar to \dace{} with the Fortran frontend as presented in Section \ref{subsec:dace}, \gtforpy{} attempts to separate the concerns of the domain scientist from those of the computational scientist and the performance engineer, even if a complete separation is not realistically possible in any framework.

In contrast to \dace{}, \gtforpy{} builds on a Python implementation. This distinguishes \gtforpy{} from the other frameworks; porting to Python represents a significant engineering effort and furthermore, making an incremental port is more difficult than for other frameworks because modules need to be interfaced with a different language. Nonetheless, we include the \gtforpy{} framework in this work because it represents yet another approach to performance portability. Moreover, in the niche case where the reference implementation is already written in Python, this porting effort disappears, giving \gtforpy{} a clear advantage over other frameworks.

In \gtforpy{} computations are expressed in a domain specific language, a subset of Python, which can be either executed natively in Python (mainly for debugging), or just-in-time (JIT) compiled to high-performance implementations for CPUs and GPUs utilizing \dace{} as an optimization framework.  On the first run of a \gtforpy{} program, the user will see a potentially significant overhead from JIT compilation workflow implemented in Python, which is avoided in subsequent runs due to caching. On the other hand, JIT compilation allows to specialize the generated executable and apply more optimizations based on the parameters passed at runtime.


The \gtforpy{} Graupel port was implemented by a computational scientist who
reproduced the original Fortran results. Performance engineers suggested
refactoring that improved the performance marginally, however, most of their
optimizations targeted \gtforpy{} and the generated \dace{} IR using custom
transformations. 
These transformations take advantage of in-place updates to memory that reduced
the number of copies necessary stemming from the functional representation of
\gtforpy{} for points that are masked-out by a \texttt{where} expression and
impacts the performance due to the sparse characteristic of the problem.
Conceptually this transformation could be implemented 
generically in the backend for all applications, but due to time constraints, this optimization was tailored to Graupel. In order to enable fair comparisons with other frameworks, we report the pure \gtforpy{} timings along with those for this specialized optimization (\texttt{\gtforpy{} opt}). 


Compared to standardized programming frameworks, the development of the \gtforpy{} DSL is currently driven by a small community.

\section{Results} \label{sec:results}

\begin{table*}[htbp]
\caption{List of HPC centers and hardware used for benchmarks in this work.}
\label{tab:hpc}
\centering
{\footnotesize
  \begin{tabular}{llrrl}
  \toprule
  System name & Organization & CPU & GPU & Notes \\
  \midrule
  Levante & DKRZ$^{\mathrm{a}}$ & AMD EPYC 7763            & Nvidia A100 80GB        &                                  \\
  Jureca  & JSC$^{\mathrm{b}}$  & AMD EPYC 7742            & Nvidia A100 40GB        & 8 NUMA domains per CPU socket    \\
  Juwels  & JSC                 & Intel Xeon Platinum 8168 &                         & GPUs not benchmarked             \\
  Beverin & CSCS$^{\mathrm{c}}$ &                          & AMD Instinct MI300      & CPUs not benchmarked             \\
  Santis  & CSCS                & Nvidia Grace             & Nvidia GH200 Hopper GPU &                                  \\
  Jupiter & JSC                 & Nvidia Grace             & Nvidia GH200 Hopper GPU & 300W CPU, 580W GPU power cap     \\
  \bottomrule
  \multicolumn{5}{l}{$^{\mathrm{a}}$Deutsches Klimarechenzentrum (German Climate Computing Centre)} \\
  \multicolumn{5}{l}{$^{\mathrm{b}}$Jülich Supercomputing Centre} \\
  \multicolumn{5}{l}{$^{\mathrm{c}}$Centro Svizzero di Calcolo Scientifico (Swiss National Supercomputing Centre)} \\
  \end{tabular}
}
\end{table*}

\begin{figure*}
  \centering
  \includegraphics[width=\linewidth]{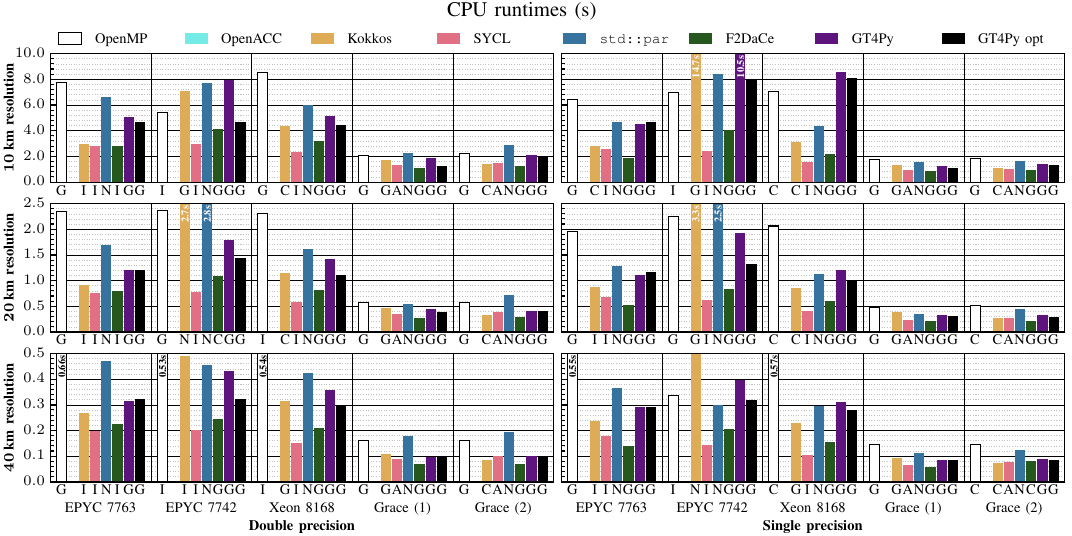}
  \caption{
  Runtimes for different implementations and grid sizes on CPU. Letters below
  the bars indicate the compiler that resulted in the fastest run:
  \underline{G}NU, \underline{C}lang/LLVM, \underline{I}ntel,
  \underline{N}vidia, \underline{A}daptive CPP. The left half shows
  double-precision runs, while the right side shows single-precision runs.
  Different grid sizes are grouped by rows; the top row is for the finest
  $10\,\mathrm{km}$ resolution, the middle row is for $20\,\mathrm{km}$
  resolution, and the bottom row is for the coarsest $40\,\mathrm{km}$
  resolution.
  The \openacc{} implementation is omitted here because no compiler
  targets it to CPU with competitive performance.
  }
  \label{fig:cpu_runtimes}
\end{figure*}

\begin{figure*}
  \centering
  \includegraphics[width=\linewidth]{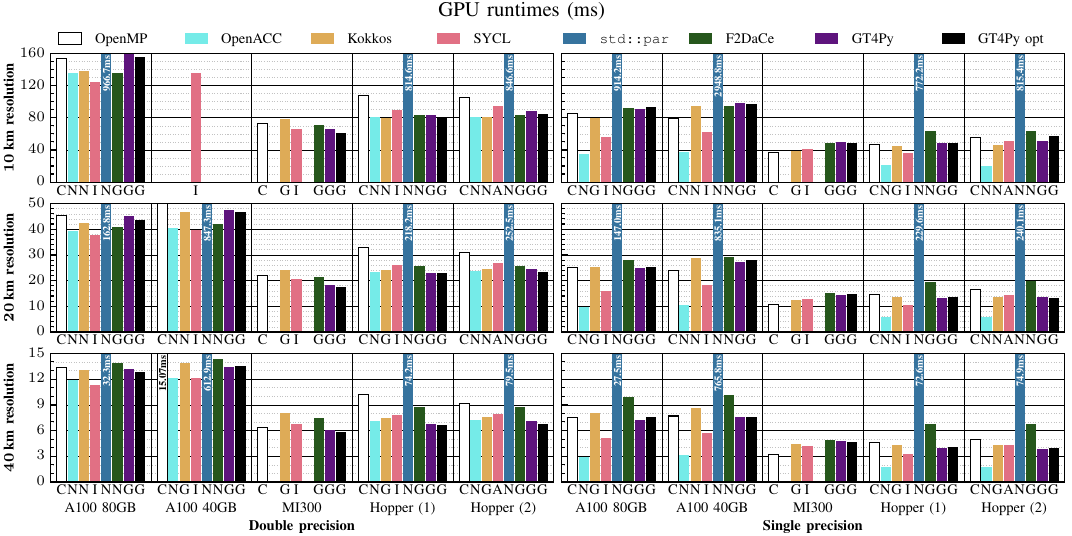}
  \caption{
  Runtimes for different implementations and grid sizes on GPU. This figure is
  structured identically to \cref{fig:cpu_runtimes}: Double-precision runs are
  shown on the left, single-precision runs on the right. The top row is for
  $10\,\mathrm{km}$, the middle for $20\,\mathrm{km}$, and the bottom row for
  $40\,\mathrm{km}$ resolution. For A100 40GB (double-precision, left half) no
  timings are shown for the $10\,\mathrm{km}$ grid because this problem size
  does not fit on the device. \sycl{} automatically shuffles memory to the
  device asynchronously, effectively circumventing this limitation. For MI300,
  \openacc{} and \stdpar{} are missing due to inadequate compiler support.
  }
  \label{fig:gpu_runtimes}
\end{figure*}


In order to understand the characteristics of each framework, we evaluate their
runtimes on several different systems with different CPUs and accelerators. On
CPU, we use the full number of cores; 64 on EPYC, 24 on Intel, and 72 on Grace.
For this reason, using our results to compare the performance of implementations
on different hardware platforms is not straightforward. Instead, we emphasize
the relative difference between implementations and how these compare between
hardware platforms and execution instances.

Each measurement reported here uses the compiler giving the best performance for
each implementation. We provide the list of viable compilers for each
combination of system, architecture, and implementation in
Appendix \ref{adx:compilers}. There, we also provide a detailed description of
compilation flags and software dependency versions.

As shown by the variations in our benchmark data, performance shifts based on a
number of factors, including not only compiler choice, framework, system, and
architecture, but also distribution, contents, and sparsity of the input data.
For this reason we introduce the term \textit{execution instance} to signify
that a specific implementation was run with a specific compiler, on a specific
system and architecture, and with a specific test size, e.g. 20km resolution.



\subsection{Performance Benchmarks}

The main experiment is designed to give each performance-portability framework
the best possible chance to perform well, while solving the same Graupel
configurations on global grids at 40 km (referred to as "R2B6" resolution), 20
km ("R2B7"), and 10 km ("R2B8") resolution across a diverse set of CPU and
accelerator platforms. The resources provided by each HPC system are listed in
\cref{tab:hpc}.
For every framework, we evaluate multiple compiler options and report results
using the compiler that achieves the best performance on each platform, rather
than imposing a uniform compiler choice across all systems. This setup is
intended to assess each framework under favorable and realistic conditions, and
to compare their ability to deliver strong performance across heterogeneous
architectures. Our results show that the best choice of compiler is not only
dependent on the framework, but also on the underlying hardware, power limits,
and problem size.

To reduce the influence of cold-start effects and capture run-to-run
variability, we execute the Graupel kernel 105 times for each measurement. We
discard the first five runs and report the median of the remaining 100
executions. We have found that the performance measurements are repeatable to
the extent that confidence intervals become effectively invisible in our plots,
so we omit them altogether. This is largely due to the fact that the lack of MPI
communication eliminates any network jitter and runs are performed in isolation
on compute nodes, so influence from the OS is minimal.

For the most part, the different frameworks perform within a factor of two of
one another for a given device. This can be observed for the GPU runtimes shown
in \cref{fig:gpu_runtimes}. Larger performance gaps appear on CPU architectures,
shown in \cref{fig:cpu_runtimes}. On CPU, the presence of multiple NUMA nodes
and various cache-coherence network layouts
can exacerbate small implementation differences, which we observe in particular
for the EPYC 7742 and Xeon CPUs, where \kokkos{}, \stdpar{}, and \gtforpy{}
perform significantly worse than their counterparts (including optimized
\gtforpy{}). EPYC 7742 has core complexes with four cores each\cite{epyc_rome}
and uses a hub-and-spoke layout for the cache-coherency layer, which is a part
of Infinity Fabric\texttrademark\cite{infinity_fabric}. In the EPYC 7763, core
complexes have eight cores each\cite{epyc_milan}, significantly reducing the
number of cache-coherence bus requests that have to be fulfilled via the slower
link to Infinity Fabric. The Intel 8168 CPU uses a mesh network interfacing all
24 cores in on NUMA domain\cite{xeon_skylake}. Grace, which also uses a mesh
network, called Scalable Coherency Fabric (SCF)\cite{nvidia_grace}, has a much
higher bisection bandwidth than the older Intel CPU, again explaining the
reduced impact of cache-coherence traffic.
\footnote{Nvidia reports a bisection bandwidth of 3.2~TB/s for SCF\cite{nvidia_grace}, while the
older Intel CPU appears to have a bisection bandwidth of about $32~\textrm{B}/\textrm{cycle} \times 2.4~\textrm{GHz} \times 5~\textrm{mesh rows} = 384~\textrm{GB}/\textrm{s}$\cite{skx_discussion,skx_uncore,cpuz_uncore}. }

Apart from a few outliers (\stdpar{} for GPU, NUMA-slowdowns on CPU), our
results show that the portability frameworks expose broadly similar
performance ceilings.
Thus, there is contention between the framework, and it is apparent that the
framework which performs best varies between execution instances, even when the
hardware of a single node is identical. For example on Hopper, \kokkos{}
performs best for the largest grid in double-precision, but is beaten by
\gtforpy{} for smaller grids. Meanwhile, on A100, \sycl{} takes a clear lead
for double precision, surpassed only by \openacc{} for single precision. Instead
of evaluating the best-performing framework, a more interesting observation here
is that the relative performance of \ftodace{} and \gtforpy{} vary between
problem sizes and precisions. For large grids, \ftodace{} wins, but \gtforpy{}
clearly performs better on the smallest $40$~km grid. Meanwhile in single
precision, \gtforpy{} wins over \ftodace{} regardless of the problem size.

Overall, \stdpar{} shows comparatively weak performance on GPU; this is caused by
the measurement problem described in \cref{subsec:stdpar}.

On CPU, where there is significantly more variance between the performance
delivered by individual implementations than on GPU, we observe that \ftodace{}
is competitive on Grace and EPYC 7763 for both precisions, but is ultimately
beaten by \sycl{} on Xeon 8186 and EPYC 7742. Unsurprisingly, \sycl{}, a
framework pushed by Intel, performs better than the other frameworks on Intel
hardware. However, the exception defines the rule: While rather lacking in
performance on EPYC and Xeon, \kokkos{} makes up for it on Jupiter's Grace CPU
(Grace (2) in the figure) to become the best implementation for small problems
in single precision. The significant difference in relative performance for the
\kokkos{} framework between Santis and Jupiter Grace CPUs is due to different
power caps set for those systems. This example clearly shows how drastic
performance variations of a single framework implementation can be triggered
by any minute difference in execution instance.

The results support our findings that no single ranking between frameworks
is possible, and a framework that performs adequately for one execution
instance may incur significant performance losses on another system or
beat all other implementations for a different problem size. Thus, we 
emphasize the need for an expressive framework that can expose different
optimizations and variants to adapt to the problem and system at hand.

\begin{figure}
  \centering
  \includegraphics[width=\linewidth]{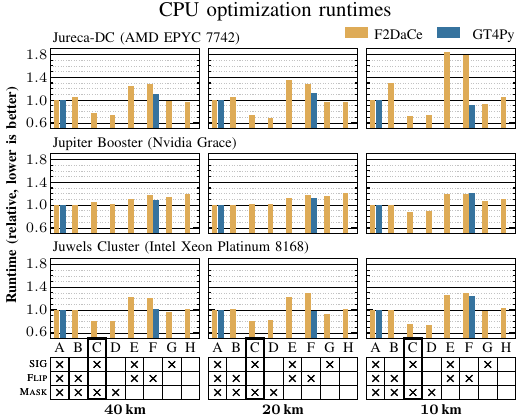}
  \caption{
  Relative runtimes for the \sig{}, \flip{} and \mask{} optimizations for all
  problem sizes on CPU. The key at the bottom indicates which of the \sig{},
  \flip{}, or \mask{} optimizations were applied. The heavy box highlights
  the configuration used by the original Fortran implementation.
  Runtimes are independently normalized so that the default implementation
  measures 1.0. Different colors show the effects of equivalent optimizations on
  different implementations.
  }
  \label{fig:sparsity_cpu}
\end{figure}

\subsection{Correctness}
An optimized implementation is considered valid only if it preserves numerical
accuracy. Correctness is evaluated using the maximum absolute difference between
double- and single-precision outputs of the original Fortran code for each
field.
Correctness checks were performed with low compiler optimization flags (-O0,
-nofma).

As we implied in the introduction (\cref{sec:intro}), testing for
bit-identical results is an unattainable undertaking, especially on GPUs. In
fact, Fortran does not impose associativity guarantees, which may be exploited
by compilers even with optimization turned off. In our Graupel case, the
unordered reduction $a+b+c$ is a concrete example of an expression that is
interpreted differently by the Nvidia and GNU compilers, leading to differences
in results. Moreover, different math libraries lead to slight numerical
deviations on CPU and GPU for various functions such as $\mathrm{log}()$ and
$\mathrm{exp}()$.
\section{Framework expressivity and engineering effort} \label{sec:expressivity}

\begin{figure}
  \centering
  \includegraphics[width=\linewidth]{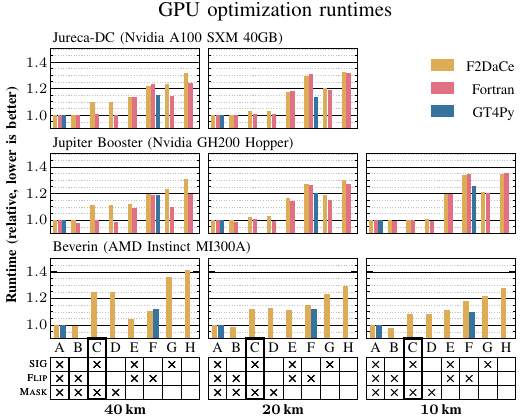}
  \caption{
  Relative runtimes for the \sig{}, \flip{} and \mask{} optimizations for all
  problem sizes on GPU. The key at the bottom indicates which of the \sig{},
  \flip{}, or \mask{} optimizations were applied. The heavy box highlights
  the configuration used by the original Fortran implementation.
  Runtimes are independently normalized so that the default implementation
  measures 1.0. Different colors show the effects of equivalent optimizations on
  different implementations. There is no data for Fortran on Beverin due to
  lack of compiler support.
  }
  \label{fig:sparsity_gpu}
\end{figure}

%
%

Our benchmark results show that all frameworks achieve good
performance on most systems. In this section, we include combinations of
optimizations that can go beyond semantic equivalence in the performance engineering space and require reasoning from the domain scientist and computational scientist as well. We focus our optimizations on how sparsity is exploited and what data access patterns are leveraged, and  analyze their impact on performance.
This leads us to consider an additional aspect when comparing performance portability frameworks, namely their ability
to express different physically equivalent implementations and the ease with which they can navigate from one to another. This is the reason why we only focus on the \ftodace{} and \gtforpy{} frameworks in this section: less engineering effort to express the different optimizations.

\subsection{Taking advantage of sparsity in Graupel} \label{subsec:sparsity}

The reference Graupel implementation makes use of two optimizations that exploit
sparsity in the fundamental physical formulation of the Graupel model. The first
optimization, which we call \mask{}, completely disables computation for any
cells where the physical properties of the air parcel do not lead to active
microphysics processes. The second optimization, which we label \sig{},
disables calculations related to \underline{s}now, \underline{i}ce, and
\underline{g}raupel when those species are only present in negligible quantities
and would thus make only negligibly small contributions to the parameterization.
As mentioned in \cref{sec:spectrum}, we discussed the impact of toggling these
optimizations, as they change the numerical output of the simulation, even if
only by a negligible amount. Domain experts confirmed that when contributions
become small in local regions of the domain, there is no difference between
counting them in or masking them out altogether. Thus enabling or disabling the
sparse optimizations as we do here does not affect the physical correctness of
the final results of the Graupel simulation:

We toggle the \mask{} and \sig{} optimizations on and off individually to
understand their impact on the performance of Graupel on different systems.

\hypothesis{On a CPU architecture, we expect each of these optimizations to
reduce runtime as they reduce the computational workload without increasing data
movement.}
\hypothesis{On a GPU architecture, we expect that these optimizations with their
associated branches increase runtime because they cause thread divergence.}

On CPU, turning off the \mask{} optimization consistently increases runtimes, as
the workload on the CPU increases: this can be seen comparing $A,B,C,D$ to
$E,F,G,H$ in \cref{fig:sparsity_cpu}. Thus, our measurements confirm our
hypothesis; \mask{} is a beneficial optimization on CPU.

Similarly, our measurements for GPU in \cref{fig:sparsity_gpu} show that the
sparsity optimization \mask{} also improves performance on GPU. The effect of
the \mask{} optimization can be seen by comparing $A,B,C,D$ to $E,F,G,H$ in
\cref{fig:sparsity_gpu}. In contrast to the CPU case, this similar result
disqualifies our hypothesis; clearly, the branch condition is not complex enough
for thread divergence to cause performance degradation.


On Nvidia GPUs, the threads that are masked out
wait to rejoin the warp after the active threads have completed their
computations. 
Even in the worst case scenario, where every wavefront suffers from thread
divergence, no additional work is done, so there is no loss of performance.
Moreover, the distribution of masked threads is typically much better than the
worst case because the computational domain is sparse (5-25\% active cells)
and exhibits high data locality due to the naturally clumped pattern of clouds.
For this reason, the \mask{} condition is homogeneous and completely on or
completely off for entire thread warps, leading to an overall reduction of work
for the GPU.

On GPU, the effects of the \sig{} when used in combination with \mask{} are
negligible. However, when \mask{} is disabled ($E$-$H$ in
\cref{fig:sparsity_gpu}), \sig{} is detrimental to performance, confirming our
hypothesis, in part.
On CPU, the effects of \sig{} are much more diverse, showing some dependence on
problem size as well as CPU architecture. On Grace, removing the \sig{}
optimization increases runtimes for the $20$ and $40\,\mathrm{km}$
grids and has close to no effect otherwise, compare $C$ to $D$ in the middle row
of  \cref{fig:sparsity_cpu}. On AMD, removing \sig{} decreases runtimes for the
$10$ and $20\,\mathrm{km}$ grids, and increases runtimes for the
$40\,\mathrm{km}$ grid.

An expressive framework is aware of minor implementation variants like \mask{}
and \sig{}, and can automatically toggle them on or off as needed for each
execution environment. Implementing these variants is still up to the
performance engineer, but transparently incorporating them into the simulation
without increasing code complexity should be handled by the framework.
\ftodace{} and \gtforpy{} take a step in this direction by making these variants
invisible to the computational and domain scientists, thus keeping code
complexity at a minimum.

Disabling either of \mask{} and \sig{} results in slight perturbations to the
numerical results of the simulation compared to the reference. However, the
reality is that the exploitation of sparsity in the Graupel reference
implementation, and thus \mask{} and \sig{} themselves, are optimizations.
Masking out non-active cells and turning off physics for locally negligible
processes results in an approximation of the true density field of microphysical
species in the atmosphere. Thus, transforming the code by disabling \mask{}
or \sig{}, which are part of the reference, results in semantically equivalent
implementations from the computational and domain scientists viewpoints. This
underscores the need for collaboration in unlocking optimization opportunities.

\subsection{Improving access patterns} \label{subsec:patterns}
Another optimization with sometimes counterintuitive effects concerns the ordering of a
nested loop that appears in the last of the kernels in Graupel. The cache
hierarchy and memory on CPU architectures generally leads to
implementations that prioritize data locality, so we hypothesize that an
arrangement with contiguous memory access in the inner loop would lead to the
best performance.
On GPU, memory loads are coalesced across warp threads, so the nested loops should
be reordered. 
For the 
nested loop in question, the inner and outer iteration statements can be
exchanged without changing the semantics of the computation, so we benchmark
both variants.
When this transformation, which we call \flip{}, is applied, we expect better
performance on GPU and worse performance on CPU.


 The \flip{} transformation does improve performance on GPU, but not to the 
expected degree. For $40\,\mathrm{km}$ and $20\,\mathrm{km}$ grid sizes, performance
does improve by up to $10\%$, but for the largest grid, the effect is negligible,
compare $C,D$ to $A,B$ in \cref{fig:sparsity_gpu}.
Once again, this is different on the AMD GPU, where the optimization introduces
almost $20\%$ speedup for $40\,\mathrm{km}$ and still reaches $10\%$ speedup for
the $10\,\mathrm{km}$ grid.


Similar to our analysis of sparse optimizations in \cref{subsec:sparsity}, more
detailed analysis shows that in reality, the scientific code introduces
additional factors which confound analytical performance predictions.
Due to the kernel size, warp threads are starved of instructions, which can
outweigh memory bandwidth bottlenecks on certain GPUs.
\subsection{Engineering effort}

The graupel microphysics component was implemented in each framework by users of
the respective framework rather than by core developers or domain experts of the
respective frameworks. This setup was chosen to better reflect a realistic
adoption scenario, in which scientific application developers must work with new
programming models and toolchains without direct access to the framework
authors. For the \openmp{}, \sycl{}, and \ftodace{} implementations, the work
was carried out by master’s students as part of their respective thesis
projects, while \openacc{} and \stdpar{} implementations are based on initial
versions created by students as part of the ISC Student Cluster Competition
(SCC).

For the \ftodace{} implementation, which relied on a prototype version of the
Fortran frontend, support from \dace{} developers was limited to assistance with
bug fixes in the tooling. Similarly, the \gtforpy{} implementation received 3-4
person months of overall support, and some new Graupel-specific optimizations
were added to the framework which is still evolving in its support for physical
parameterizations.
Beyond technical assistance, the objective was for each implementation to be
developed as independently as possible by its assigned developer.


Although a perfectly level playing field cannot be achieved in a study of this
scale, particularly given the limited sample size and the differing maturity
levels of the frameworks, all implementations were carried out under a common
guiding principle: to obtain a best-effort solution within a reasonable
development period.  The intention was therefore not to compare idealized expert
implementations, but rather to assess what can be achieved in practice by
capable users working under realistic time constraints.
Moreover, we can find one execution instance for each framework in which it
outperforms all other implementations: \openmp{} is best for MI300
single-precision but is outperformed by \gtforpy{} on MI300 double-precision. On
other GPUs, \openacc{} outpaces the competitors for single-precision. \kokkos{}
outperforms the others for single-precision on Grace, while \sycl{} takes the
lead on the Intel Xeon CPU. Finally, \ftodace{} leads the pack for
double-precision on Grace. Thus, each framework is best at least once,
regardless of differences in developer skill and invested time. The only
exception is \stdpar{}, which performs poorly across the board, mostly due to
measurement problems (\cref{subsec:stdpar}).

While porting the applications to each framework took roughly similar amounts of
work in each case, we observed \ftodace{} and \gtforpy{} to be at an advantage
when introducing additional variants as described in the introduction to
\cref{subsec:sparsity}. This is because the \dace{}-based frameworks decouple
fine-grained optimizations from implementation, enabling developers to make
changes at a high level that are immediately reflected in the optimized
executable.

For other frameworks, the parallelization scheme is coupled to the code structure,
so implementing changes like \flip{} would require re-associating the relevant
directives (\openmp{}, \openacc{}) or abstractions (\kokkos{}, \sycl{}, \stdpar{})
with the correct loop body. However, \dace{} and \gtforpy{} are only robust against
simple changes like \sig{}, \flip{}, and \mask{}; large scale restructuring will
still cause brittle failures.

In this way, \ftodace{} and
\gtforpy{} are more expressive in decoupling fine-grained optimizations from
high-level changes, allowing the computational scientist to evaluate the
performance impact of high-level changes more rapidly. This is because in most
cases, \dace{} can simply carry over the fine-grained optimizations introduced
by the performance engineer without requiring any manual intervention.

%
%
%
%

\subsection{Framework expressiveness}

\researchquestion{The challenge in predicting the actual performance impact across architectures of these reasonably simple optimizations highlights the need for performance portability frameworks to quickly and easily express such concepts, as only having the flexibility to produce multiple variants and test them can lead to finding the solution with the best performance.}


Thus, in an attempt to achieve excellent performance on every system, one would
need to pick the most suitable from a number of implementations on every system.
In many cases, this approach requires an infeasible engineering
effort to develop implementations with all frameworks.

Instead, a framework could allow the performance engineer to suggest, but not
enforce possible performance optimizations, while exposing as many semantic
constraints as are necessary to maintain correctness, but \textit{not more.}
The goal is to
enable the performance engineer to be as \textit{expressive} as possible when
preparing the scientific application for the performance framework in order to
enable the framework to best optimize for a target architecture.

\kokkos{} and \sycl{} are expressive in that they let performance engineers encode semantic constraints through a range of abstract parallelization strategies, albeit at the cost of requiring domain and computational scientists to understand the application in those terms. \gtforpy{} and \ftodace{} similarly enable richer semantic information to reach the automatic optimization stage through their data-centric internal representations, while allowing scientific developers to continue working from a simpler source formulation. This greater expressiveness typically comes with substantially more complex internal machinery—as seen in \kokkos{}’ use of modern C++ compile-time mechanisms and, even more so, for \gtforpy{} and \ftodace{}, which effectively operate as compilers in their own right—but that complexity is precisely what can simplify the overall development process and improve collaboration between domain scientists, computational scientists, and performance engineers.

\section{Related Work} \label{sec:related}

The tools covered in our work represent the state of the art
when it comes to performance portability,
given their ability to generate code for a majority of CPUs
and GPUs used routinely in HPC.\footnote{
That is, with the exception of \stdpar{}, which is young compared to the other
frameworks and is limited by compiler support, enjoying no support on AMD
GPUs.}
However, other tools, approaches, and explorations of 
performance portability exist. We list and discuss appropriate representatives here. 

Python provides Numba~\cite{numba}, an LLVM Just-in-Time compiler which
transforms Numpy code into machine code for better performance. During the
compilation process, Numba resolves pointer indirection and lowers a subset of
high-level Python and Numpy features into efficient loops with performance
comparable to other low-level compiled languages. In particular, Numba is able
to apply aggressive transformations by limiting the scope of supported features
and deviating from Python semantics.~\cite{numba} Superficially, this is similar
to the \gtforpy{} DSL which uses \dace{} \sdfg{}s as an IR instead of Numba's
LLVM MLIR, and which has been specialized and optimized for stencil-based
computations. Similar to our \gtforpy{} implementation, using Numba would still
require a complete re-write of the Graupel code in Python. Given that Numba
provides less opportunities for influencing the optimization process, we expect
performance to be at most comparable with \gtforpy{}. In lieu of DSL features, a
Numba port may also require more lines of code to achieve the same optimized
stencil operations which are exposed as a language feature of \gtforpy{}.


The Alpaka~\cite{Zenker_2016} and RAJA~\cite{raja} frameworks provide a C++11
abstraction layer for hierarchical parallelism. In this regard, they are similar
to \kokkos{} and \sycl{}, though they require no specialized compiler, unlike
the latter. For these reasons, we expect performance and implementation effort
in RAJA and Alpaka to be comparable to \kokkos{} and \sycl{}. Added
functionality, like loop chain abstractions~\cite{raja-loops} in RAJA, can offer
additional performance benefits.


The hardware vendors Nvidia and AMD offer the CUDA and HIP programming languages
that can be used to implement kernels for their respective accelerators. These
languages offer the most flexibility when it comes to extracting performance,
but the engineering effort required to write an optimized program grows
proportionally. Moreover, CUDA and HIP prioritize their respective
accelerators\footnote{Although HIP provides utilities to convert HIP from CUDA
and even run HIP code on Nvidia devices.} and do not produce portable software
outside a narrow range of devices. For these reasons, we do not compare our
results against a CUDA/HIP baseline. Nonetheless, the \dace{} backend for GPU
generates CUDA code, so this can be interpreted as a baseline.

Comparable initiatives to our own exist to evaluate and compare performance
portability frameworks. Harrell et al.\ \cite{harrell18} develop a tool to
facilitate the calculation and collection of performance portability data. Using
their tool, they explore the effectiveness of multiple performance portability
frameworks including \openmp{}, \kokkos{}, and CUDA. Additionally, they
characterize development effort using a subjective task load
index~\cite{hart88}.
Artigues et al.\ \cite{artigues20} reimplement a Particle-In-Cell (PIC) kernel
in \openmp{}, \openacc{}, CUDA, \kokkos{}, and RAJA targeting CPU and GPU,
finally turning to an in-depth discussion of the usability and performance of
\kokkos{} and RAJA.
Marowka~\cite{marowka22} evaluates 324 case studies of \openacc{}, \openmp{},
\kokkos{}, and RAJA based on a novel performance portability metric.
Deakin et al.~\cite{babelstream} develop BabelStream, a framework to evaluate
the maximum memory bandwidth that can be attained using different programming
models, including \kokkos{}, \openmp{}, \openacc{}, and \sycl{}.
Daniel et al.\ \cite{daniel19} benchmark various applications ported to
\kokkos{} and \openacc{} on CPU and on GPU, highlighting that the optimal
framework depends on problem size, among other factors. This supports our
results, driving home the idea that multiple implementations are required
to extract optimal performance from an arbitrary system.


Focusing on the effort needed to port codes to various frameworks,
Heroux et al.~\cite{heroux24} offer a comprehensive retrospective on portability
frameworks used to port codes to GPU within the Exascale Computing Project
(ECP). In a similar vein, Dubey et al.~\cite{dubey21} give a perspective on the
adoption of frameworks like \openmp{}, \kokkos{}, and RAJA in ECP projects. The
authors highlight the importance of close collaboration between domain
scientists and computer science experts. Many publications inhabiting specific
domains, like Roccon et al.~\cite{roccon26} for computational fluid dynamics,
Bauer et al.~\cite{bauer21} for earth-system science, or Abdulah et
al.~\cite{abdulah25} for statistical computing recognize the need for
portability but focus on domain-specific challenges and requirements. To the
best of our knowledge, the current literature lacks a discussion of framework
expressivity and how it ties into development effort at all levels.

\section{Conclusion} \label{sec:conclusion}


The clear takeaway from our experiments is that portable applications must
flexibly exchange optimizations in order to adapt to different hardware
configurations and problem configurations to extract the highest possible
performance. What remains to be investigated is how to achieve this level of
expressivity, which is not yet offered by any existing framework. In addition to
this, porting codes to any framework efficiently is an open research problem,
especially when the framework in question is specially designed to make use of
multiple implementation variants for some kernels. In a similar vein, Holmen et
al.\ \cite{holmen19} propose an approach for adopting portability frameworks in
a piecewise fashion. Building on these ideas, further research is required to
work out the details of porting existing code to a framework without
re-implementing and testing many sections of the algorithm in different variants
to achieve sufficient expressivity, regardless of human work hours or LLM tokens
spent.

Our evaluation shows that, despite relying on markedly different programming
models, abstraction mechanisms, and optimization strategies, all considered
performance-portability frameworks are ultimately capable of approaching a
similar performance ceiling for the Graupel kernel on modern systems. This
suggests that no single framework holds a fundamental monopoly on achievable
performance. Rather, the main differences lie in how that performance is
obtained: through different balances of manual intervention, compiler support,
abstraction design, and framework-specific optimization opportunities.

At the same time, our results indicate that these differences have important
consequences for the development process itself. The effort required to obtain
high performance, and the way this effort is distributed among domain
scientists, computational scientists, and performance engineers, varies
substantially across frameworks. Some approaches demand that semantic
restructuring and optimization intent be made explicit early in the
implementation, increasing the burden on those expressing the scientific kernel,
while others defer more of this work to intermediate representations, compiler
tech\-no\-lo\-gy, or downstream transformation stages. Consequently, the choice
of framework should not be assessed solely in terms of runtime performance, but
also in terms of how it shapes collaboration, maintainability, and the division
of expertise required to bring scientific applications efficiently onto new
architectures.

\section*{Acknowledgment}
%
We gratefully acknowledge Hussein Harake and the Swiss National
Supercomputing Centre (CSCS) for providing access to computing resources
via the RACKlette student cluster, which was instrumental to this work.

The authors acknowledge the ICON partnership — including the German Weather
Service (DWD), the Max Planck Institute for Meteorology, the German Climate
Computing Center (DKRZ), the Karlsruhe Institute of Technology (KIT),
MeteoSwiss, and ETH Zurich — for their continued development and improvement of
the ICON model. The work of many members of this partnership over the years has
laid the foundation for the results presented in this study. The further
development of ICON for the efficient use of exascale HPC systems was provided
by the German Federal Ministry of Research, Technology and Space (BMFTR) funded
project WarmWorld, under grant number 01LK2203.This work used resources of the
Deutsches Klimarechenzentrum (DKRZ) granted by its Scientific Steering Committee
(WLA) under project ID bk1341. This work was supported by the
\mbox{WeatherGenerator} project (grant agreement No. 101187947). This work was
supported by the Swiss National Science Foundation (SNSF) through the National
Centre of Competence in Research (NCCR CLIM+ -- Climate Extremes and Society).

\bibliographystyle{IEEEtran}
\bibliography{bib-general,bib-software,bib-dace,bib-prec}

\begin{appendices}

\section{Compilers}\label{adx:compilers}

\begin{figure}
  \centering
  \includegraphics[width=\linewidth]{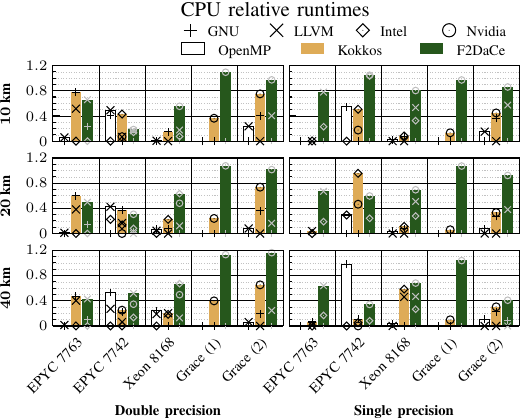}
  \caption{Comparison of compiler-influenced performance variance for CPU. Bars
  show the relative difference between fastest and slowest compilers: 
  ($(\mathrm{max}-\mathrm{min})/\mathrm{min} = \mathrm{max}/\mathrm{min}-1$).
  Only \openmp{}, \kokkos{}, and \ftodace{} have a free choice of compiler on
  CPU (see \cref{tab:compilers}). Given that tallest bars show a relative
  difference of about $1$, this plot confirms that overall, the best results
  have about half the runtime of the slowest.}
  \label{fig:cpu_ccs}
\end{figure}

\begin{figure}
  \centering
  \includegraphics[width=\linewidth]{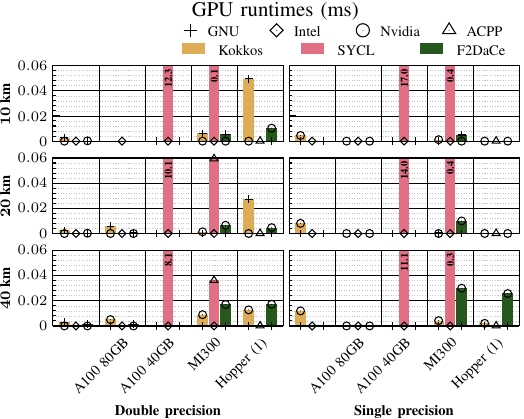}
  \caption{Comparison of compiler-influenced performance variance for GPU. Bars
  show the relative difference between fastest and slowest compilers: 
  ($(\mathrm{max}-\mathrm{min})/\mathrm{min} = \mathrm{max}/\mathrm{min}-1$).
  Only \kokkos{}, \sycl{}, and \ftodace{} have a free choice of compiler on GPU
  (see \cref{tab:compilers}). The relative difference between compilers is
  minimal compared to to CPU (\cref{fig:cpu_ccs}), staying mostly below $5\%$.
  The exception is formed by ACPP, which performs much more poorly than Intel.}
  \label{fig:gpu_ccs}
\end{figure}

\begin{table*}[htbp]
\newcommand{\x}{$\times$}
\newcommand{\s}{$\mathbin{\mathop{\scalebox{1.5}{$\ast$}}}$}
\newcommand{\nC}[1]{\multicolumn{1}{c}{#1}}
\newcommand{\yC}[1]{\multicolumn{1}{c|}{#1}}
\caption{Table showing compilers tested for each implementation on each system.
The abbreviations G, C, I, N, A correspond to the GCC, LLVM, Intel, Nvidia, and
Adaptive CPP compilers.}
\label{tab:compilers}
\centering
\begin{tabular}{lccccc|ccccc|ccccc|ccccc|ccccc}
\toprule
\multicolumn{1}{c}{}            & \multicolumn{5}{c}{x86\_64}
                                & \multicolumn{5}{c}{aarch64}
                                & \multicolumn{5}{c}{sm\_80}
                                & \multicolumn{5}{c}{sm\_90}
                                & \multicolumn{5}{c}{gfx942} \\
\multirow{2}{*}{Framework}      & \multicolumn{5}{c}{EPYC, Xeon}
                                & \multicolumn{5}{c}{Grace}
                                & \multicolumn{5}{c}{A100}
                                & \multicolumn{5}{c}{Hopper}
                                & \multicolumn{5}{c}{MI300A} \\
\cmidrule(lr){2-6} \cmidrule(lr){7-11} \cmidrule(lr){12-16} \cmidrule(lr){17-21} \cmidrule(lr){22-26}
           &G&C&I&N&\nC{A}  &G&C&I&N&\nC{A}  &G&C&I&N&\nC{A}  &G&C&I&N&\nC{A}  &G&C&I&N&\nC{A}  \\ \midrule \rC
\yC{\openmp}    &\x&\x&\x&  &    &\x&\x&  &  &    &  &\x&  &  &    &  &\x&  &  &    &  &\x&  &  &    \\
\yC{\openacc}   &  &  &  &  &    &  &  &  &  &    &  &  &  &\x&    &  &  &  &\x&    &  &  &  &  &    \\ \rC
\yC{\kokkos}    &\x&\x&\x&\x&    &\x&\x&  &\x&    &\x&  &  &\x&    &\x&  &  &\x&    &\x&  &  &  &    \\
\yC{\sycl}      &  &  &\x&  &    &  &  &  &  &\x  &  &  &\x&  &\x  &  &  &\s&  &\x  &  &  &\x&  &\x  \\ \rC
\yC{\texttt{std::par}}    &  &  &  &\x&    &  &  &  &\x&    &  &  &  &\x&    &  &  &  &\x&    &  &  &  &  &    \\
\yC{\ftodace}   &\x&\x&\x&\x&    &\x&\x&  &\x&    &\x&  &  &\x&    &\x&  &  &\x&    &\x&  &  &  &    \\ \rC
\yC{\gtforpy}   &\x&  &  &  &    &\x&  &  &  &    &\x&  &  &  &    &\x&  &  &  &    &\x&  &  &  &    \\
\bottomrule
\multicolumn{26}{l}{\s\ On ALPS/Santis, the legacy Intel DPC++ compiler is still available. At the time of writing, this compiler and the accompanying PTX plugin are} \\
\multicolumn{26}{l}{\phantom{\s\ }neither supported nor served, making a new install on Jupiter Booster impossible.}
\end{tabular}
\end{table*}

\begin{table*}[htbp]
\newcommand{\tb}[1]{\textbf{#1}}
\caption{Compiler flags used for each compiler and implementation. Flags printed in bold are not available on aarch64.}
\label{tab:flags}
\centering
{\footnotesize
  \begin{tabular}{cll}
  \toprule
  \multicolumn{2}{c}{Compiler} & \multicolumn{1}{c}{Flags}                                     \\ \hC
  \multicolumn{2}{c}{} & \textcolor{white}{Performance}                                        \\ \rC
  G & GCC    & \texttt{-O3 -ffast-math}                                                        \\
  C & LLVM   & \texttt{-O3 -ffast-math -ffp-contract=on -std=c++17 -march=native -fuse-ld=lld} \\ \rC
  I & Intel  & \texttt{-O3 -std=c++17 \tb{-fma -march=core-avx2 -mtune=core-avx2}}             \\
  N & Nvidia & \texttt{-O3 -std=c++17 -fast}                                                   \\ \rC
  A & ACPP   & \texttt{-O3 -ffast-math -ffp-contract=on -march=native}                         \\
  \hC
  \multicolumn{2}{c}{} & \textcolor{white}{Correctness}                                        \\ \rC
  G & GCC    & \texttt{-g -O0 \tb{-mpc-64}}                                                    \\
  C & LLVM   & \texttt{-g -O0 -fuse-ld=lld}                                                    \\ \rC
  I & Intel  & \texttt{-g -O0 -fp-model=precise \tb{-m64}}                                     \\
  N & Nvidia & \texttt{-g -O0 -nofma \tb{-Mnofprelaxed -Msingle}}                              \\ \rC
  A & ACPP   & \texttt{-g -O0}                                                                 \\
  \bottomrule
  \end{tabular}
}
\end{table*}

\begin{table*}
\caption{Summary of clusters used for benchmarks in the paper. The table
includes: system name, CPU hardware, GPU hardware, compiler versions for GCC,
LLVM, OneAPI, and NVHPC. N/A is given for ALPS/Beverin CPU Hardware and for
Juwels Cluster GPU Hardware because those combinations were not used.
Furthermore, the Intel OneAPI compiler suite is not available for ARM-based
hardware (ALPS/Beverin, ALPS/Santis, Jupiter Booster). AdaptiveCpp on
ALPS/Beverin uses LLVM~18.1.0.}
\label{tab:systems}
\centering
{\footnotesize
  \begin{tabular}{lllccccc}
  \toprule
  \multicolumn{3}{c}{}
  & \multicolumn{5}{c}{Compiler versions} \\
  \multicolumn{1}{c}{System Name}
  & \multicolumn{1}{c}{CPU Hardware}
  & \multicolumn{1}{c}{GPU Hardware}
  & \multicolumn{1}{c}{GCC}
  & \multicolumn{1}{c}{LLVM}
  & \multicolumn{1}{c}{OneAPI}
  & \multicolumn{1}{c}{NVHPC}
  & \multicolumn{1}{c}{AdaptiveCpp} \\ \midrule \rC
  Levante         & AMD EPYC 7763            & Nvidia A100 SXM4 80GB & 12.3.0 & 17.0.4  & 2023.2.1  & 24.7     & N/A            \\
  Jureca DC       & AMD EPYC 7742            & Nvidia A100 SXM 40GB  & 13.3.0 & 18.1.8  & 2024.2.0  & 24.9     & 25.10          \\ \rC
  ALPS/Beverin    & N/A                      & AMD Instinct MI300    & 13.4.0 & $\phantom{^\mathrm{a}}$20.0.0$^\mathrm{a}$  & N/A       & N/A      & 25.10          \\
  ALPS/Santis     & Nvidia Grace             & Nvidia GH200 Hopper   & 13.2.0 & N/A     & N/A       & 25.1     & 25.10          \\ \rC
  Juwels Cluster  & Intel Xeon Platinum 8168 & N/A                   & 13.3.0 & 20.1.8  & 2025.2.0  & 24.9     & N/A            \\
  Jupiter Booster & Nvidia Grace             & Nvidia GH200 Hopper   & 13.3.0 & 18.1.8  & N/A       & 24.9     & 25.10          \\
  \bottomrule
  \multicolumn{8}{l}{$^{\mathrm{a}}$18.1.0 for AdaptiveCpp} \\
  \end{tabular}
}
\end{table*}

\begin{table}
\caption{Versions for software dependencies of our toolchain. The last column
indicates CUDA versions for every cluster except ALPS/Beverin, where the ROCM
version is generally 7.1.0, but 6.3.3 for AdaptiveCpp only.}
\label{tab:ver}
\centering
\begin{tabular}{lcccc}
\toprule
                & \multicolumn{4}{c}{Software versions}               \\
System name     & CMake      & NetCDF C++      & CDO      & CUDA/ROCm \\ \rC
\midrule
Levante         & 2.26.3     & 4.3.1           & 2.2.2    & 12.5      \\
Jureca-DC       & 3.30.3     & 4.3.1           & 2.6.0    & 12.6      \\ \rC
ALPS/Beverin    & 3.28.3     & 4.3.1           & N/A      & $\phantom{\mathrm{a}}$7.1.0$^\mathrm{a}$     \\
ALPS/Santis     & 3.27.9     & 4.3.1           & 2.4.0    & 12.6      \\ \rC
Juwels Cluster  & 3.30.3     & 4.3.1           & 2.6.0    & N/A       \\
Jupiter Booster & 3.29.3     & 4.3.1           & 2.6.0    & 12.6      \\
\bottomrule
\multicolumn{5}{l}{$^{\mathrm{a}}$6.3.3 for AdaptiveCpp}
\end{tabular}
\end{table}

\cref{tab:compilers} shows the compiler-implementation combinations that were
considered for each hardware platform. The compilation flags for each compiler
are collected in \cref{tab:flags}, while \cref{tab:systems} shows the version
numbers for compilers on each system. For completeness, \cref{tab:ver} gives
the version numbers for software dependencies.

\cref{fig:cpu_ccs,fig:gpu_ccs} visualize the differences in performance
influenced by compiler choice. Finally, \cref{tab:best} list the cases in which
a particular framework is best-in-class for our experiment. This information
is extracted from \cref{fig:cpu_runtimes,fig:gpu_runtimes}.


\begin{table}
\caption{In which execution instances does a framework offer best-in-class
performance? This table lists execution instances in which a particular
implementation performs better than the others. Notably, there is at least one
case for each implementation, except \stdpar{}. Furthermore, the lists are not
comprehensive, e.g. \sycl{} is also best-in-class for double precision on EPYC
7742.}
\label{tab:best}
\centering
\begin{tabular}{ll}
\toprule
Implementation  & Best-in-class performance for: \\ \rC
\midrule
\openmp{}       & MI300, single precision, all resolutions       \\
\openacc{}      & A100, Hopper single precision, all resolutions \\ \rC
\kokkos{}       & Grace (2), single precision, R2B6              \\
\sycl{}         & Xeon 8168, single precision, all resolutions   \\ \rC
\stdpar{}       & \textit{N/A, see \cref{subsec:stdpar}}         \\
\ftodace{}      & Grace, double precision, all resolutions       \\ \rC
\gtforpy{}      & MI300, double precision, all resolutions       \\
\bottomrule
\end{tabular}
\end{table}
\end{appendices}

\end{document}